# Lorentz Transformation as a Complementary Time-Dependent Coordinate Transformation. The Physical Foundation of Einstein's Special Relativity

A.C.V. Ceapa
alex_ceapa@yahoo.com

We here deduce Lorentz transformation (LT) as a member of a class of time-dependent coordinate transformations, complementary to those already known as spatial translations and rotations. This exercise validates the principle of physical determination of equations within special relativity theory (SRT), in accordance with the derivation of the LT in Einstein's original paper on relativity. This validation is possible because our LT deduction also discloses the real physics warranting Einstein's manipulation of several equations in that paper, thus proving the correctness of his derivation of the LT. The essential role of the revelation in the act of science also results. Far from being an arbitrary dogmatic construction, SRT appears to be a revealed dogmatic theory that can be turned into a truly physical one like operational theory. Radically new technological applications of relativistic quantum theory then result.

## l. Introduction

We here obtain and apply to SRT time-dependent coordinate transformations that are complementary to those already known as spatial translations and rotations. We name the new coordinate transformations 'complementary time-dependent coordinate transformations', and denote them by CT's. The CT's are derived by projecting onto coordinates axes the radius vectors of geometrical points in inertial spaces when traced by physical signals (the physical signals determine the direction and magnitude of the radius vectors at the moment of their projection). Inertial physical space is an assembly of geometrical points at rest with respect to each other, aimed by uniform rectilinear motion as a whole [1]. Unlike physical inertial spaces, empty sapace is the assembly of geometrical points endowed with no motion at all. The need to trace radius vectors with physical signals arises when they change systematically over time, in both direction and magnitude relative to inertial observers.

Since recording physical signals is a measurement procedure, our method of investigation is an operational method. The source of the physical signals is attached to the origin of the inertial coordinate system (CS) of an observer. Any inertial CS is an assembly of three straight lines orthogonally crossing at a point, that moves uniformly and rectilinearly as a whole [1]. The coordinate axes are determined by the bodies, fixed relative to each other, that constitute a reference frame (RF). As all their geometrical points are those of their axes, the CS's are embodied in physical spaces. The source's emission is isotropic. It takes place when the observer's CS coincides with a CS at rest in the physical space to which the geometrical points belong. Only one of the emitted signals reaches such a point. The origin of this signal is designated by a point in empty space, and those of the two CS's, by points in the physical spaces to which they belong. The first is a point at absolute rest, while the last two are points aimed by uniform rectilinear motions.

The three origins, and the geometrical point the radius of which was traced by signal, are joined together by a mathematical relationship which, in reduced form, associates abstract CS's at absolute rest (CSAR's) with the two inertial CS's (Sects 2.1, 2.2). The axes of an abstract CS are not determined by the physical bodies of a RF. So, an abstract CSAR does not presume the existence in Nature of an RF at absolute rest. An abstract CSAR cannot violate the principle of relativity. So we deduce the CT's by assuming the concepts of absolute rest, absolute motion and absolute speed, as well as the concept of absolute time (by identical clocks running at the same rate, independently of their speed). There results a deep insight on **i)** the 'relativistic' law for the composition of parallel velocities (Sect. 2.1), **ii)** the measurement of absolute speeds by inertial observers (Sect. 2.2.1), and **iii)** the light-speed principle (Sect. 2.2.2). By adding travel times as scalar quantities in Sect. 3 (*i.e.*, by an equation obtained dividing a geometrical equation by a signal's speed), the mathematical relationship reduces to a CT. The CT obtained for light signals is identical to the standard LT (Sect. 4). This result validates our assumed concepts.

Einstein was the only physicist fairly close to deducing the LT as a CT [2]. However, he emitted the signal at a time when the two inertial CS's did not coincide, and a priori ignored any CSAR. Consequently, he couldnot realize that, by that signal, he actually traced a radius vector. So, he failed in concluding that the tracing of radius vectors by physical signals was the objective reality underlying his decisions (pointed out in Sect. 5) to manipulate several equations in order to obtain the LT in [2]. So much the more, he couldnot conclude that this objective reality validated the principle of the physical determination of equations in SRT[1], so providing the physical grounds of SRT. Two years later [3], and then ever after, Einstein and his followers opted for using purely mathematical derivations of the LT. SRT was developed without the derivation of the LT in [2]. Relativity handbooks and papers based on them, as well as science-fiction productions exploiting their misleading physical predictions, have flourished ever since.

Neither Einstein nor his followers worried about the disastrous impact the chosen dogmatic formulation of SRT would have upon human knowledge and progress in physics and technology: His *a fortiori* formulation of the light-speed principle and the concepts of spacetime and time dilation, have broken the logical relation between the typical concepts of space and time, motion and rest, and absolute and relative. The need to abolish the original significance of those concepts could never be proved. Nevertheless, the breaking has violently penetrated the human consciousness for a century now.

The absence of the principle of physical determination of equations in SRT has led to relativistic quantum theories deprived of both physical foundations and a large amount of information on the subquantum structure of matter. For more than 65 years nobody became aware of this fact. So much more that testing and exploiting this information needed to develop new techniques. Instead of developing such techniques, physicists have asked for more and more powerful accelerators of quantum particles. Due to the absent information, they have not succeeded in understanding and systematizing the current data obtained by colliding ultra-relativistic particles. Groups of experimenters who obtained brand new results have associated each of them to separate mathematical explanatory models that have hidden their common nature. Consequently, they couldnot refine their techniques in order to achieve real advanced technologies. Other radically new technologies remained beyond imagination.

The relativism of the last century (unfairly claiming support from Einstein's SRT) continues successfully the dissolution of both scientific and common knowledge, with major consequences upon economy, society, *etc*. At least strangely, leaders in science and technology policy have opted to ignore this dramatic state of the affairs for at least the next fifty years. The trend to describe the whole physical universe, including the microcosm, in terms of geometry of a claimed physical spacetime and its quantum nature dominates [4-6] , against its striking failure [7]. Our disclosure in this paper of the operational nature of the Minkowskian space-time does not support a physical spacetime. Moreover, by validating the principle of the physical determination of equations in SRT, we provided entirely new information on the subquantum structure of matter [8]. This information is essential to achieve a unified theory of elementary-particle interaction. What is responsible for the crisis of modern physics (see Appendix 3) is not SRT, but the deliberate disregard of Einstein's 1905 [2] derivation of the LT.

The physical foundations of SRT were implicit in Sect. I.3 of [2] but SRT was built without the derivation of the LT in [2]. Section 6 of the present paper demonstrates the correctness of the derivation of

1 It every term of the underlying equations of any physical theory has incorporated an explicit physical significance. This is what we call the principle of the physical determination of equations. This principle was basic to develop classical physics. However, it was never defined in physics textbooks, and its special importance for the advancement of physics was never pointed out.

LT in [2]. But Einstein failed to give such a rationale for his mathematical decisions; he just imposed them authoritatively. This proves that those decisions were simply 'revealed' to him (see Appendix 4). Experimental results confirm the essential predictions of quantum relativistic theories. So SRT was a revealed dogmatic theory that can now be turned into a physical one as an operational theory (Sect. 7). SRT is not a deliberate dogmatic construction, as most relativity handbooks persuasively insinuate. So scientists should brake both the atheistic mentality (beneficiary of a formidable logistics) and the mentality that revealed knowledge cannot be turned into rational knowledge by a rationale, to substantially improve their creative performance. These results are of ultimate importance at a time when a large amount of scientific information is missed by discarding the role of revelation in the discovery act (act of science: the birth of any new idea, or set of coupled ideas, contributing to the advancement of science). Our operational method is confirmed by the derivation of the vector LT (Appendix 1) and the proof that collinear LT's form a group (Appendix 2A). The LT is validated by the operational proof (Appendix 2B) that non-collinear LT's form a group without requiring rotations of inertial CS's to this end [9]. The objective reality warranting the manipulations of equations that led to the LT in [2], provided by our operational method, validates not only the derivation of the LT in [2] but also the concepts we initially assumed to deduce the CT's in Einstein's SRT. Conclusions are drawn in Sect. 8. This paper is the up to date of [10].

## 2. Abstract CSAR's

In the following subsections we give evidence for abstract CSAR's associated to inertial CS's 'at rest' and abstract CSAR's that 'professional' inertial observers (professionals) associate to their own inertial CS's. A 'professional' is an observer who is *a priori* trained to represent graphically both real and fictitious relative motions.

### 2.1 Abstract CSAR's Associated to CS's 'at Rest'

Consider the diagrams in Fig. 1, with arrows temporarily ignored. In the first diagram, the CS k is moving with constant speed $v$ along the positive common $x', x$ axis relative to a hypothetical CSAR K. In the second diagram, k moves with the same speed relative to $K_1$, but k and $K_1$ are carried by an inertial physical space of speed $w$. CS k coincided with both K and $K_1$ at $t = 0$. P($x'$) is a fixed point in k. At time $t$ the second diagram differs from the first one in that everything is shifted right by a distance $wt$. The Galilean transformation

$$x' = x - vt \tag{1}$$

is predicted by both diagrams. This fact 'entitled' observers to name their inertial CS's 'at rest', and to treat them as CSAR's.

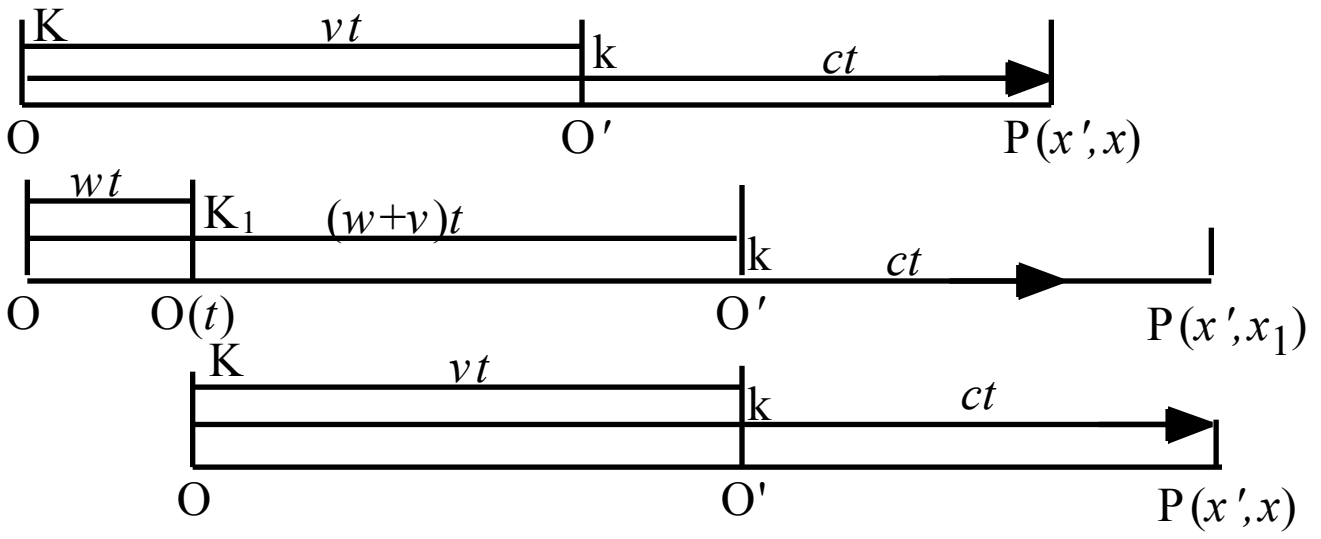

Figure 1.

Among all possible physical signals, let us here select light signals to reveal the deep connection of our CSAR's and CT's with SRT. Let the arrows on Fig. 1 stand for the light signal tracing the radius vector of P( $x'$). At time $t = 0$, this signal and the origin of k leave the origin of K, $K_1$, respectively, moving along the $x', x$ axes with speeds $c, v$. At time $t$, they reach, respectively, P and O' in the first diagram, and we get Eq. (1) with $x = ct$. Also at time $t$, the path of the signal in the second diagram is $ct$, but both the origin of $K_1$ and P are shifted right to O(t) and P($x', x_1$) for the distance $wt$. At time $t_1 = t + wt / c$ the light signal will reach P($x', x_1$), but in the time $wt/c$, P($x'$) moved from P($x', x_1$) to P($x', x_2$) in the diagram of Fig. 2. At time $t_2 = t_1 + (w+v)wt/c^2$, the light signal will reach P($x', x_2$), while k, $K_1$ and P($x'$) moved further to right by $(w+v)w^2t/c^2$, and $(w+v)^2wt/c^2$, respectively. So that, the time $t_f$, at which k and $K_1$ will reach positions denoted respectively by k($t_f$) and $K_1$($t_f$), and the light signal P($x'$) at P($x', x_f$), tracing its radius vector relative to O, is given by

$$t_f = t + wt/c + (w+v)wt/c^2 + (w+v)^2 wt/c^3 + ...$$
$$= t + wt/c + (w+v)wt/c(c - w - v)$$

where the sum of an infinite geometric series of common ratio $(w+v)/c < 1$ was taken into account. At time $t_f$, the radius vectors of P($x'$) and of the origin of k, respectively, are located at

$$x_f = ct_f = ct + wt + (w+v)wt/(c - w - v)$$

and

$$x_{O'} = (v+w)t_f = vt + wt + (w+v)wt/(c - w - v)$$

So $x_f - x_{O'}$ reduces to Eq. (1) by removing the line segments OO($t$) $= wt$ and $P(x', x_1)P(x', x_f) = (w+v)wt/(c - w - v)$ covered by the light signal and the origin of k, in accord with the second diagram in Fig. 1. The third diagram in Fig. 1 follows. By that the radius vector of the geometrical point P($x'$) is traced by the signal in time $t$, this diagram associates the abstract CSAR K to the observer's inertial CS $K_1$. Therefore, the very graphical and mathematical description of the uniform rectilinear motion of any object relative to an inertial observer is done with respect to the CSAR associated to his inertial CS. The 'relative' speed appears to be an absolute quantity (an absolute quantity is one defined with respect to a CSAR).

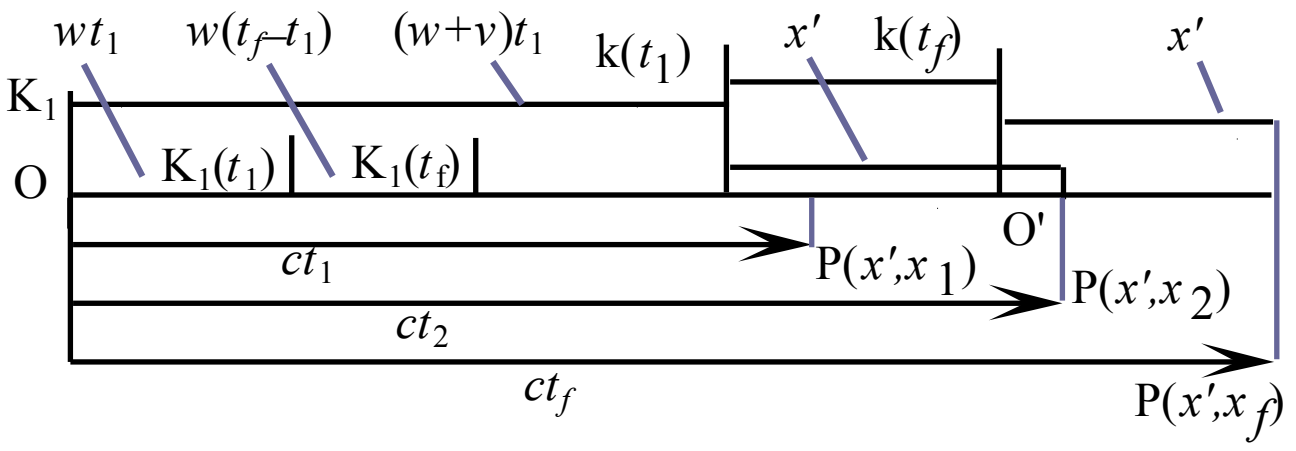

Figure 2.

Consider now the diagrams in Fig. 3. The CSAR K is that above associated to $K_1$. The $k_A$, k and K coincide at $t_0 = 0$. Just at $t_0 = 0$, $k_A$, k and a light signal, tracing the radius vector of P fixed in k, leave the origin O of K. They move uniformly along the common $x', x'', x$ axis with absolute speeds $v, w$ and $c$, respectively. At time $t$, their origins and the tip of the signal reach, respectively, the points $O'_A$($vt$), O'($wt$) and Q($ct$) in the upper diagram. By diagrams like the last two in Fig. 1, with $K_1$, K changed to $k_A$, $K_A$, we turn the motion of k relative to $k_A$ to one relative to the CSAR $K_A$ associated to the inertial $k_A$. To this end, the light signal and the origin of k must continue their motion an additional time $vt/c$, until reaching P and O'[$w(t + vt/c)$], respectively. Since $O'_A$P was traveled by the signal in time $t$, the bottom diagram in Fig. 1 is regained as the second one in Fig. 3, where O'($t$), O'($t'$) stand for the origin of k relative to $O'_A$ at times $t, t'$, respectively. For a speed $u$ of k relative to $K_A$, this diagram predicts the relationship $ut' = (w - v)t$ at the time $t' = t - wvt/c^2$ and, by simplification, the equation

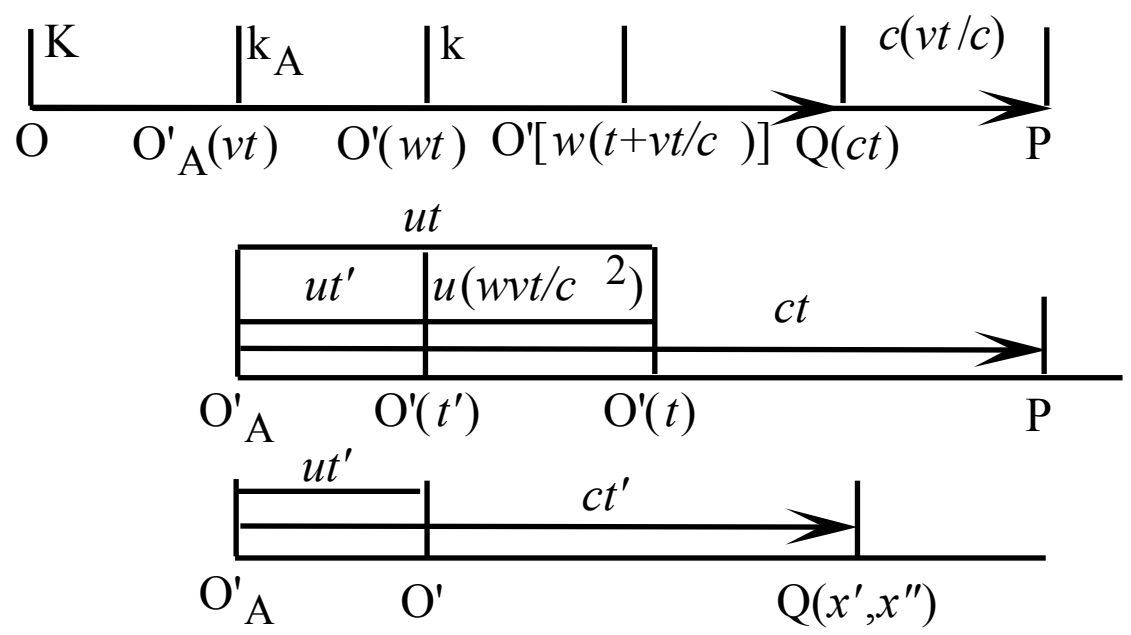

Figure 3.

$$u = (w - v) / (1 - wv / c^2) \quad (2)$$

Hence, the 'relativistic' Eq. (2) is specific to a theory in which the radius vectors are traced by light signals. The first two diagrams in Fig. 3 predict the time-dependent coordinate transformation

$$x' = x - vt, \; t' = t - vx / c^2 \quad (3)$$

by the additional equation $x = wt$ (expressing the identification of Q with the origin O′ of k), which assures the independence of Eqs. (3). This coordinate transformation relates the translatory motions of constant speeds $u, w$ of an object (the origin of k) relative to the CSAR's (K, $K_A$) associated to inertial CS's, ($K_1$, $k_A$). Since Eqs. (3) and the equations

$$x'' = x - wt, \; t'' = t - wx / c^2$$

also predicted by the first diagram, give rise to the equations

$$x'' = x' - ut', \; t'' = t' - ux'/c^2$$

predicted by the last diagram, the coordinate transformation (3) forms a group.

## 2.2. Abstract CSAR's Associated by Professionals to Their Own Inertial CS's

Professionals at rest with respect to the origin of k, and professionals at rest with respect to $K_1$ in Fig. 1, can associate CSAR's (Ξ) to any k by reflecting at point P($x'$) fixed in k the light signal tracing its radius vector, as depicted in the diagrams in Fig. 4. The first because, as a point of space, hence at absolute rest, the origin $O'_o$ of the signal defines the origin O of K, the last in view of the third diagram in Fig. 1. They get the equations

$$x' = x - vt, \; x' = vt_1 + ct_1 \quad (4)$$

having as solutions

$$t = x'/(c - v), \; t_1 = x'/(c-v). \quad (5)$$

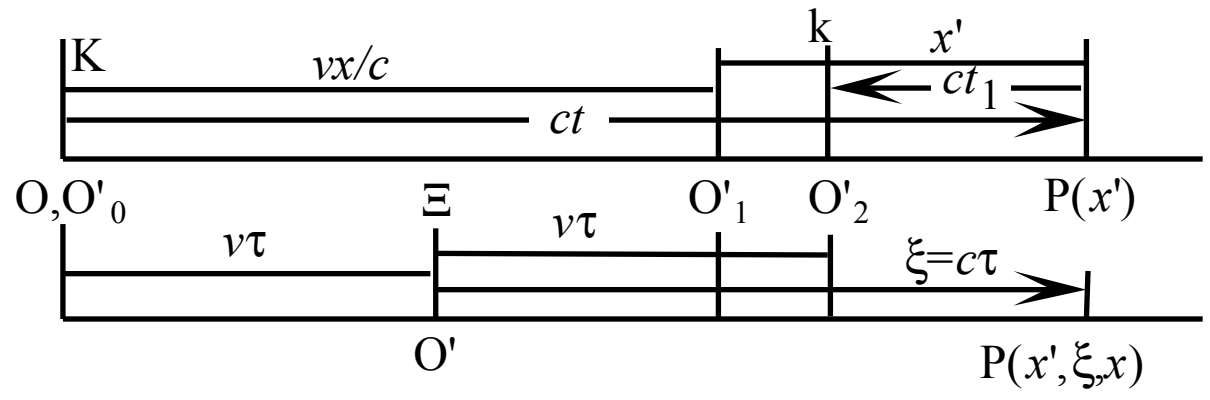

Figure 4.

Thus defining

$$\tau = (t + t_1) / 2, \; \xi = c\tau \quad (6)$$

they obtain equations $\tau = \beta^2 x'/c$, $\xi=\beta^2 x'$, and implicitly

$$O'_oO'_2/2=c(t-t_1)/2=v\tau \quad x = ct = \xi + v\tau \tag{7}$$

Since $x'$ is the abscissa of a point P fixed in k, it is constant. The quantities $\tau$ and $\xi$ are also constants. Therefore, the point O' of abscissa $v\tau$ is a fixed point in K. Since $\xi$ gives the position of P relative to O′, the last of Eqs. (7) defines O′ as the origin of a CSAR $\Xi$ associated to the inertial CS k. As depicted in the second diagram in Fig. 4, $\Xi$ is parallel to k and K, having in common the $x', \xi, x$ axis. The radius vector of P relative to $\Xi$, $\xi$ is traced by a light signal in the time $\tau$ of $\Xi$. By Eqs. (5), (6) and (1), and adding equations $\eta = y$, $\zeta = z$, he gets

$$\xi = \beta^2(x - vt),\ \eta = \beta y,\ \zeta = \beta z,\ \tau = \beta^2(t - vx/c^2) \tag{8}$$

where $\beta = 1/\sqrt{1 - v^2/c^2}$, which connect coordinates of P relative to the CSAR's $\Xi$, K.

### 2.2.1. Determination of Absolute Speeds

Examine the ability of professional inertial observers to determine experimentally absolute speeds. The upper diagram in Fig. 4 is useful in this aim. The resulting Eqs. (4) have as solutions the absolute speeds

$$v = x'(t - t_1)/2tt_1 \text{ and } c = x'(t + t_1)/2tt_1 \tag{9}$$

Therefore, such observers can actually determine their absolute speeds, as well as that of light, by measuring travel times of light signals traveling to and fro along arbitrary directions. To this end, each of them has to emit to P($x'$) at time $t = 0$ a light signal which origin, as a point of space (hence at absolute rest), defines that of an 'unseen' CSAR K, coinciding with his k. When the measured times $t, t_1$ are equal, $v = 0$ and the light speed in empty space is just $x'/t$. The experiment must be repeated along other directions until $v$ in (9) takes a maximum value. That value defines the absolute speed of k, while the path of the suitable light signal determines its direction of motion.

That $c \pm v$ are not true light speeds, we show in view of the second diagram in Fig. 1. First presume that k is attached to an object $M_2$ moving rectilinearly with constant speed $v_2$ on the plane surface of another object $M_1$ (having $K_1$ attached), along the constant speed $v_1$ of $M_1$ or oppositely. The relative speeds $v_1 \pm v_2$ are true physical quantities: They appear as true speeds of $M_2$ in both its kinetic energy and linear momentum. Imagine that $M_1$, $M_2$ are moving rectilinearly, uniformly, simultaneously and independently in vacuum at speeds $V_1$ and $\pm V_2$, respectively. This time the relative speeds $V_1 \pm V_2$ are not true physical quantities: They do not appear as true speeds of an object. They manifest physically by transfer of linear momentum when the two bodies collide each other. The last is the case with the quantities $c \pm v$, appearing by the factorization mathematically required to resolve Eqs. (4) in terms of $t, t_1$, respectively: the simultaneous parallel motions, that of the light signal traveling in empty space between $O'_o$ and P($x'$), and that of k from $O'_o$ to $O'_1$, are fully independent.

### 2.2.2 Light-Speed Principle

Einstein's assertion [11] that "The totality of physical phenomena is of such a character that it gives no basis for the introduction of the concept of 'absolute motion'" is contradicted by the result we just obtained. We see that the simultaneous and independent motion of the line segment O′P in Fig. 4 along the $x$ axis as a part of k alters the equality of the paths of the light signal from the origin of k to P($x'$) ($O'_oP$) and back to the origin of k ($PO'_2$), stipulated by the light-speed principle. It does not matter that isolated inertial observers are not aware of this alteration. It is their assumed lack of knowledge on the relative motion responsible for this fact. The experiment just proposed for measuring absolute speeds proves it: For $O'_oP$ to equate $PO'_2$, the light signal should have been made of elastic balls rolling *on* a surface embodying the $x'x$ axis from the origin of k to P($x'$) and back to the origin of k, which is not the case. Therefore, the light-speed principle was stated in relation to the CSAR associated to the inertial CS of the observer in Sect. 2.1. The rigor of SRT was assured just by this intuitive hidden formulation of the light-speed principle, which tacitly imposed the abstract CSAR to the inertial observers. In view of this result, as well as of those obtained in Sects. 2.1, 2.2 and 2.2.1, Einstein's queer aversion for 'absolute motion' and CSAR's was baseless and misleading.

## 3. 'Graphical' Addition of Travel Times

Consider a sequence of collinear line segments $OA_1$, $A_1A_2$,…, $A_{n-1}A_n$ in empty space, and denote

$$OA_n = OA_1 + A_1A_2 + \ldots + A_{n-1}A_n. \tag{10}$$

Because the time in which a light signal travels any line segment is the difference between the times indicated by synchronous clocks located at its endpoints at the arrival of that signal [in our case $t$(O), $t$ ($A_1$), … , $t$ ($A_n$)], we always have

$$t\,(OA_n) = t\,(OA_1) + t\,(A_1A_2) + \ldots + t\,(A_{n-1}A_n) \tag{11}$$

with $t\,(OA_n) = t(A_n) - t(O) = OA_n/\,c$, $t(OA_1) = t(A_1) - t(O) = OA_1/\,c$, $t(A_1A_2) = t(A_2) - t(A_1) = A_1A_2/\,c$, … , $t(A_{n-1}A_n) = t(A_n) - t(A_{n-1}) = A_{n-1}A_n/\,c$.

The choice of collinear light signals in SRT has hidden the case of the collinear line segments which depend on travel times of non-collinear light signals, like those tracing the radius vectors OQ, O'Q in the diagram in Fig. 5, with k and K in Sect. 2.1. The collinear line segments OO', O'P and OP are covered respectively with speeds $v$, $c\cos\alpha$ and $c\cos\theta$ by the origin of k and the projections onto the common $x', x$ axis of the tips of the light signals tracing OQ, O'Q. Therefore they depend on the travel times $t^*$ and O'Q $/c$. Evidently, this prevents us from getting a time equation like (11) by simply dividing equation OO'+O'P=OP by $c$. In order to get such an equation, we need to express OP, OO' and O'P in terms of the travel time of one and the same light signal. This means that we need to relate them geometrically to the path of such a signal (O'$P_1$ in Fig. 5). We name *time-axis* the direction orthogonal to $\mathbf{v}$. By applying the Pythagorean theorem to the right triangle O$P_1$O', we have

$$t^* = \beta t_- \tag{12}$$

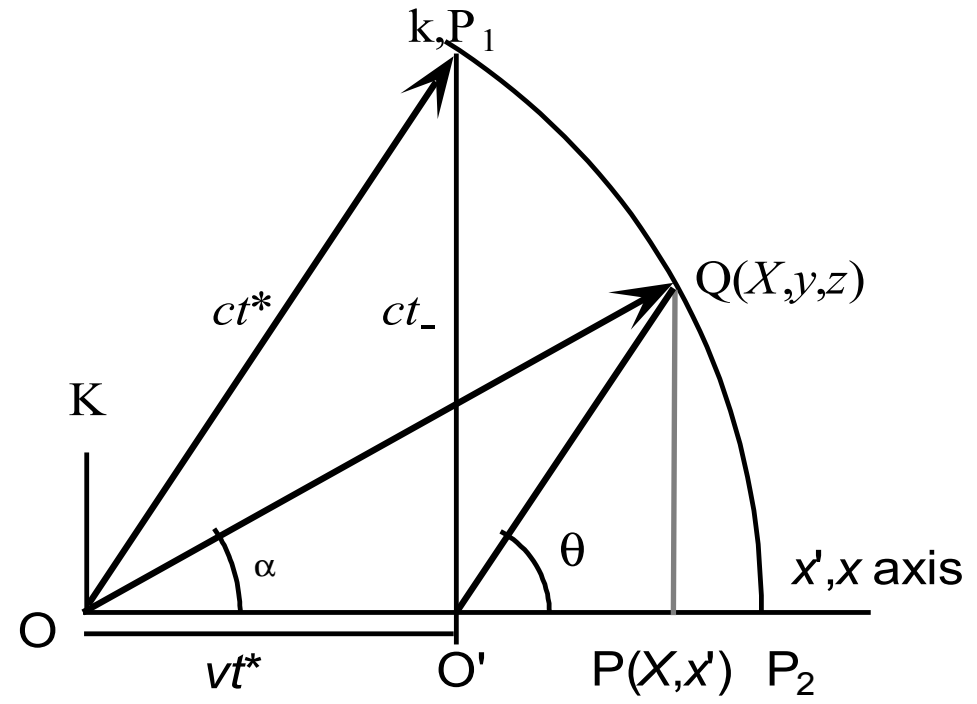

Figure 5.

Laying O'O and OP on the time-axis is straightforward. Similarly expressing O'P as the path of a light signal fails, in which case we must identify different geometry avoiding the dependence of O'P on O'Q/ $c$.

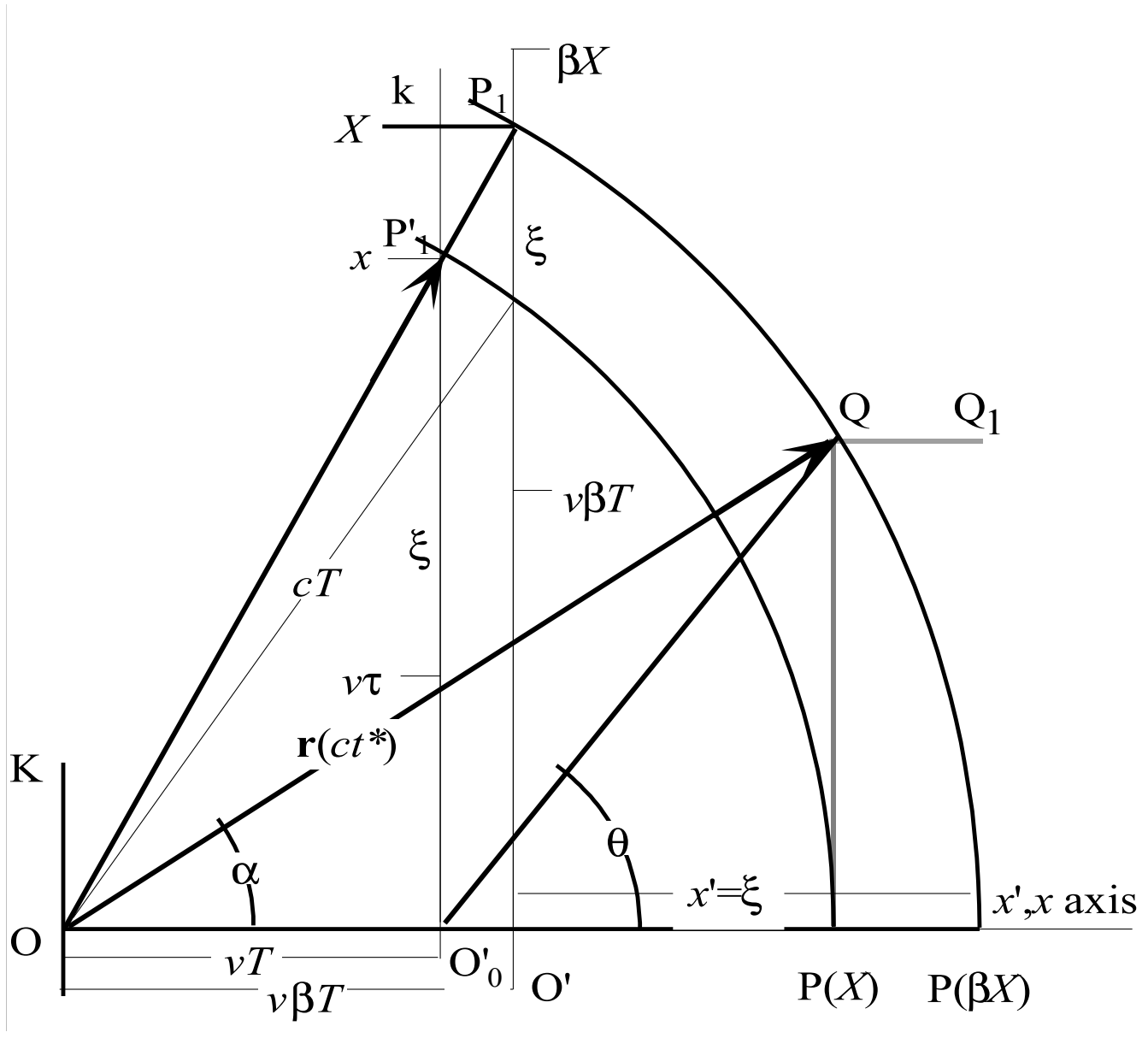

Figure 6.

Consider the diagram in Fig. 6, also with k and K in Sect. 2.1. Q, $Q_1$, and P( $X$), P( $\beta X$) as their projections, are fixed points relative to k. At time $t = 0$, the origin of k and the light signal traveling to P( $X$ ) leave the origin O of the CSAR K. At time $T$ [$(r/c)$ cosα], they reach, respectively, $O'_o$ and P( $X$). We lay the bottom diagram in Fig. 4 at $O'_o$ on the time-axis $O'_oP'_1$ which means that we refer the motion of k to the CSAR Ξ For the reason leading to (12), from the right triangle $OP'_1O'_o$ we have

$$T = \beta t, \; X = cT = \beta ct = \beta x, \; OO'_o = vT = v\beta t \quad (13)$$

By Eqs. (7), (6), and (13) we further determine $\xi$ and $\tau$ in terms of $X$ and $T$. We get

$$\xi = \beta(X - vT), \; \eta = y, \; \zeta = z, \; \tau = \beta(T - vX/c^2). \quad (14)$$

Thus by passing from Q to the geometrical point $Q_1$, we get rid of the dependence of the abscissa of P on the time $O'_oQ/c$. The abscissa of $Q_1$ relative to K is $\beta$ times that of Q. It is $\xi$ with respect to both k and Ξ: Since $\xi$ is traveled by a light signal in time $\tau$, the abscissa of $Q_1$ relative to k is also traveled in time $\tau$.

Therefore, a time equation like that given by (11) follows immediately along the $x', x$ axis, dividing by $c$ the equation OO'+O'P( $\beta X$)=OP( $\beta X$). So, for any geometrical point P( $x'.x$), we have the set of equations

$$x' = \beta(x - vt), \; y' = y, \; z' = z, \; t' = \beta(t - vx/c^2) \quad (15)$$

$\beta x$ in (15) is the Cartesian coordinate of a geometrical point associated to P( $x'.x$) in consequence of the graphical addition of travel times as scalar quantities, like travel time $\beta t$ is a Newtonian time, while $\beta vt$ also a Cartesian coordinate.

## 4. The Standard LT as a CT: The General Form of the CT's

For Eqs. (15) to express a coordinate transformation, we must brake the equivalence of the first and the last of them. To this end, consider the Q's (implicitly their projections P) in Fig. 6 to move relative to the CS k, which is also in uniform translatory motion relative to K. Identifying P with the origin of CS k, we are in the case pointed out in the last paragraph of Sect. 2.1. Therefore, we pass from a description of the motion of Q relative to the inertial CS k to one with respect to a CSAR $K_A$ associated to k just as it was associated to $k_A$ in Sect. 2.1. By a diagram analogous to the last one in Fig. 3 and by the additional equation $x = wt$ analogous to that associated to Eqs. (3), we break the equivalence of the first and the fourth of Eqs. (15). The terms $\beta x$, $\beta vt$ and $\beta t$ in Eqs. (15) keep the same meaning of Cartesian coordinates and Newtonian time deduced in Sect. 3. Thus, with the additional equation $x = wt$, Eqs. (15) deduced by tracing radius vectors by light signals represent the standard LT like a CT. The new derivation of the LT predicts neither physical length contraction nor physical time dilation (against [12], and in accord with [13]). The increased lifetimes of the relativistic particles with respect to identical rest particles originate in relativistic mass as internal coupling constant [8]. The new derivation of the LT does not support maintaining paradoxes (mathematical speculations on the precariously stated physical grounds of SRT). E.g., the paradox in [14] does not involve that the LT should always connect *infinitesimals* instead of finite coordinates. For an inertial observer attached to the origin of the CS S′ in the diagram in [14], who traces by light signals radius vectors with respect to the origin of S′, does not merely exist any paradox. As any CT is defined by the speed of the physical signal tracing radius vectors (let it be $\upsilon$), the general form of the CT's is given by Eqs. (15) with $c$ changed to $\upsilon$ in Eqs. (15).

## 5. Einstein's Derivation of LT in [2]: The CSAR in Einstein's SRT

Einstein defined clocks working in synchrony at points O′, P 'of space' (see Sect. I.1 of [2]), *i.e*, at absolute rest, by the equation

$$\tau_0 + \tau'_0 = 2\tau_P \tag{16}$$

where $\tau_0$, $\tau'_0$ and $\tau_P$ are, respectively, times associated to the emission/arrival of a light signal at O′, and its reflection at P. He deduced the LT in [2] in view of a Gedanken experiment depicted in the diagram in Fig. 7, and by imposing three mathematical decisions with no physical justification. The first decision was to extrapolate the validity of Eq. (16) to clocks at rest at O' and P in the inertial CS k. Then, from the diagram in Fig. 7 (with k and K in Sect. 2.1), which differs from the upper one in Fig. 4 in that the signal was emitted at time $t$ when k and K did not coincide, he defined and calculated $\tau$ (like time of k) in terms of the time $t$ of K, and the coordinates of a point having P as projection. He inserted the times $\tau_0 = \tau(0,0,0,t)$ associated to the emission of a light signal at $O'_1$, $\tau_P=\tau[x',0,0,t+x'/(c-v)]$ associated to reflection at P, and $\tau'_0 = \tau[0,0,0,t+x'/(c-v)+x'/(c+v)]$ associated to its arrival at $O'_2$, where $O'_o$ to $O'_2$ are successive positions of the origin O′ of k along the common $x',x$ axis, in Eq. (16) and obtained for infinitesimally small $x'$ the differential equation

$$\partial\tau/\partial x' + [v/(c^2 - v^2)]\partial\tau/\partial t = 0 .$$

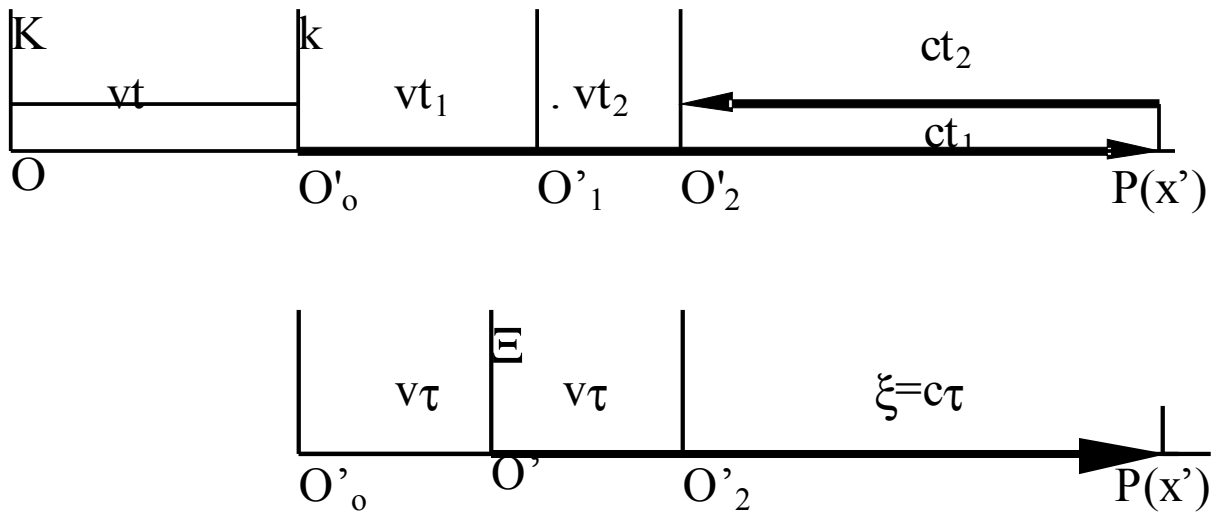

Figure 7.

Integrating this equation, he obtained

$$\tau = a[t - vx'/(c^2 - v^2)] \tag{17}$$

with $a = \phi(v) = 1$ (see Sect. I.3 of [2] for calculation details), and put

$$\zeta = c\tau . \tag{18}$$

Accepting Eq. (1), Eqs. (17) , (18) predicted a set of equations linear in $\beta^2$ identical with Eqs. (8). Einstein's s second decision[2] was to drop the square of $\beta$ in Eqs. (8). It is true of Eqs. (8), as well as of their counterparts linear in $\beta$, that the last one is the time-equivalent of the first one. Einstein did not point out this equivalence, or the way to break this equivalence for turning the linear equations in $\beta$ into the LT. His third decision was to add the equation $x = wt$ to the linear equation in $\beta$, in order to deduce the law of addition of parallel speeds (in Section I.5 in [2]).

[2] Prokhovnik claimed in [15] that Einstein had included a $\beta$ factor in Eqs. (8) in the function $\Psi(v) = \beta\phi(v)$. However,

The physical grounds for decisions 1 to 3, and hence their correctness, we disclose in the next Section.

## 6. The Physics Warranting Einstein's Mathematical Decisions to Deduce the LT in [2]

Since the diagrams in Fig. 7 are those in Fig. 4 shifted right by a distance $vt$, they predict, respectively, equations identical with Eqs. (4) to (7), with $t, t_1$ changed to $t_1, t_2$, $t_1$=$\tau_P$-$\tau_0$, $t_2$=$\tau'_0$-$\tau_P$, and $\tau_0$, $\tau_P$, $\tau'_0$ in Sect. 5. Therefore, by his first decision Einstein tacitly associated both abstract CSAR's and 'professionals' to the inertial CS's in SRT. As pointed out in Sect. 2.2.2, his light-speed principle was actually stated in relation to CSAR's. The coordinates $\xi, \eta, \zeta$ were defined with respect to the CSAR $\Xi$. What the inertial synchronous clocks located at O' and P in the bottom diagram of Fig. 7 measure (by Eq. (16) and by O'P+PO'= $2c\tau$) is the time $\tau$ of $\Xi$. Behind Einstein's second decision, *i.e.* to drop the square of $\beta$ in Eqs. (8), lies the graphical addition of travel times like scalar quantities, investigated in Sect. 3. Without the diagram in Fig. 6 for points out of $x'$ axis, Einstein failed in understanding $\beta x$ and $\beta t$ as a Cartesian coordinate and a Newtonian time. Thus $\beta x$ and $\beta t$ were conceived, respectively, as a coordinate and a time multiplied by a mysterious factor $\beta$, which led to the famous length contraction and time dilation. The last paragraph in Sect. 2.1 proves that the role of the equation $x = wt$, imposed by Einstein's third decision, was to remove the equivalence of the first and the fourth of Eqs. (15) to turn them into a coordinate transformation.

These physical grounds for Einstein's firm mathematical decisions prove the correctness of his derivation of the LT in [2]. Faced with his early tacit ignorance of that derivation of the LT, this correctness proves that those decisions were 'revealed' to him. These conclusions remained undisclosed for about a century just because the light signal in Fig. 7 was emitted at a time when the CS's k, $K_1$ did not coincide, and Einstein failed to see the operational nature of his method.

## 7. Operational Theories

Physical theories embodying CT's are operational theories. Physical quantities defined in CS's moving uniformly and rectilinearly with respect to inertial observers depend on both time and spatial coordinates. So they are expressed relative to their CSAR's by CT's [16], [17]. The CT

$$x' = x - ct,\; t' = t - x/c \tag{19}$$

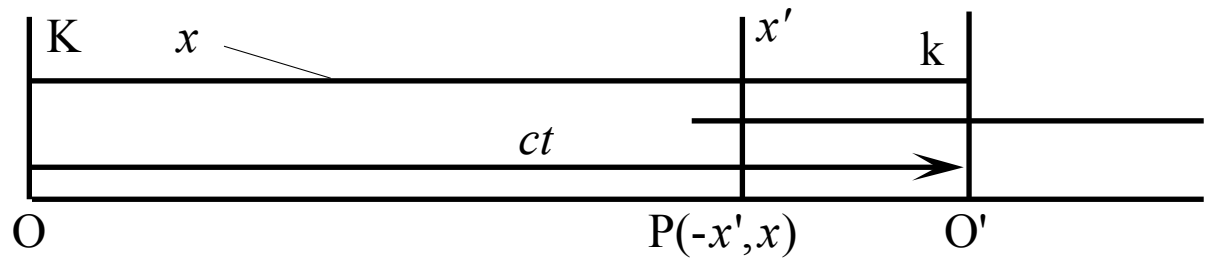

Figure 8.

is obtained for $v = c$ from the particular CT (3), associated to the diagram in Fig. 8. It makes the electromagnetic and general relativity theories operational theories by predicting the dependence on $t - x/c$ of the retarded potentials, of the potentials defining plane waves traveling along the $x$ axis, and of the quantities $f$ and $\xi^i$ ($i = 0$ to $3$) that appear in their gauge transformations [18] as well. This time dependence was imposed by the coincidence with experiment and the purely mathematical Lorentz condition in electromagnetism, and by the gravitational counterpart of the latter [18] (alternatively the transverse-traceless conditions [6]) in general relativity theory. Thus, predicting the time-dependence of

---

there is no function $\Psi(v)$ in [2]. Moreover, it is evident that Einstein did not include a $\beta$ factor in $\phi(v)$, given that the $\phi(v)$ appearing in the equations linear in $\beta$ that he finally wrote in [2] is just that which he formerly associated with Eq. (17).

their potentials and the relationships connecting them by (19), Einstein's equations of the gravitational fields in vacuum, *viz*.

$$R_{ik} = 0$$

where $R_{ik}$ is the Ricci tensor, define plane gravitational waves as physical entities.

## 8. Conclusions

**1)** The LT belongs to the class of CT's for the first time defined in [16]. Unlike the ordinary time-dependent coordinate transformations, the LT can be written *only when* the radius vector of a geometrical point in an inertial space is physically traced by a light signal. The mixture of spatial coordinates and times in the LT, and the Minkowski space-time originates just in this tracing. Therefore, space-time has an operational nature. It means three-dimensional space plus time, and has no connection with the spacetime claimed to be a physical entity causing physical effects [4-6]: The spacetime has no physical grounding and no physical effect. With this remark, SRT contributes to a unified theory of elementary-particle interaction.
**2)** The operational derivation of the vector LT in Appendix 1 and the operational proof in Appendix 2A that the *collinear* LT's form a group validate our operational method, while the operational proof in Appendix 2B that the *non-collinear LT's form a group,* without requiring rotations of inertial CS's [9] in this aim, validates the LT itself. The operational law for the composition of non-parallel velocities established in Appendix 2B does not predict any Thomas precession [9], [19], in accord with the experimental facts [20].
**3)** Since by the graphical addition of travel times like scalar quantities $\beta x$ and $\beta vt$ are Cartesian coordinates, and $\beta t$ a Newtonian time, the LT predicts no true Fitzgerald-Lorentz contraction, as well as no true time dilation, thus the meaning of the Newtonian concepts of space and time is keept unaltered in SRT, in deep agreement with everyday experience and common sense. The larger life-times of the relativistic particles, unnaturally connected to the 'time-dilation', have a different origin [8].
**4)** No operational theory embodying CT's can challenge SRT. This because $c$ in SRT is also the speed of real and virtual photons implied in quantum and subquantum processes.
**5)** The operational identification of the abstract CSAR in the SRT and of the terms in the LT as Cartesian coordinates and Newtonian time validates the classical principle of the physical determination of equations in SRT. Involving that $m_0c^2$ and $\beta m_0c^2$ are energies of particles at absolute rest and moving with constant speed $v$ in relation to a CSAR (with $\beta$ only coinciding with the $\beta$ of the LT as long as an observer moving with absolute speed $v$ also sees $\beta m_0c^2$ as the energy of a particle at absolute rest), they also validate this principle in relativistic quantum mechanics, thus providing genuine information on the subquantum structure of matter that the statistical interpretation of quantum mechanics could never provide, particularly concerning the subquantum energy, essential to found radically new technologies [8]. Unfortunately, for about a century now, the way to disclose and apply such information was firmly forbidden by editorial policies as part of physics policy.
**6)** All inertial observers are equal to one another in SRT, but, by enabling them to refer physical quantities measured in their RF's to quantities defined in CSAR's by LT, the SRT is a *theory of the absolute*, having nothing in common with the almighty misleading relativism governing the 20th and now 21st century. This conclusion is supported by limits that our results in Sects 2.1 and 2.2.1 set to Einstein's statement [10] that "The name 'theory of relativity' is connected with the fact that motion from the point of view of possible experience always appears as the relative motion of one object with respect to another. Motion is never observable as 'motion with respect to space' or, as it has been expressed, as 'absolute motion'." Thus: **i)** Even though motion always appears, from the point of view of experience, as the relative motion of one object with respect to another, this happened only because the inertial CS attached to the latter object was named 'at rest' and erroneously treated as a CSAR; **ii)** That such motion is always observable as absolute motion by any inertial observer was proved in Sect. 2.2.1.

## Appendix 1: The Vector LT

Consider the diagram in Fig. 9. The CS k moves rectilinearly with constant speed $v$ relative to the CSAR K along the direction $\hat{\mathbf{v}} = \mathbf{v} / v$. A light signal traveling OP in time $T$ is used, just like in Sect. 3 above (O'P' playing the role of time-axis), to remove the dependence of OP and O'P on $t^*$ and O'Q/ $c$, respectively, by passing from Q and O' to $Q_1$ and $O'_1$ with $OP_1 = \beta OP$ and $OO'_1 = \beta OO'$. From the right triangles $O'_1Q_1P_1$ and OQP we have $\mathbf{r}' = \mathbf{Q_1P_1} + \mathbf{O'_1P_1}$ with $\mathbf{Q_1P_1} = \mathbf{r} - (\mathbf{r}\cdot\hat{\mathbf{v}})\hat{\mathbf{v}}$ and $\mathbf{O'_1P_1} = \mathbf{OP_1} - \mathbf{OO'_1} = \beta[(\mathbf{r}\cdot\hat{\mathbf{v}})\hat{\mathbf{v}} - \mathbf{v}T]$, that by noting, $t' = \mathbf{r}'\cdot\hat{\mathbf{v}}/c$ and $T = \mathbf{r}\cdot\hat{\mathbf{v}}/c$, provides the vector LT as

$$\mathbf{r}' = \mathbf{r} - (\mathbf{r}\cdot\hat{\mathbf{v}})\hat{\mathbf{v}} + \beta[(\mathbf{r}\cdot\hat{\mathbf{v}})\hat{\mathbf{v}} - \mathbf{v}T], \; t' = \beta[T - \mathbf{r}\cdot\mathbf{v}/c^2] \qquad (20)$$

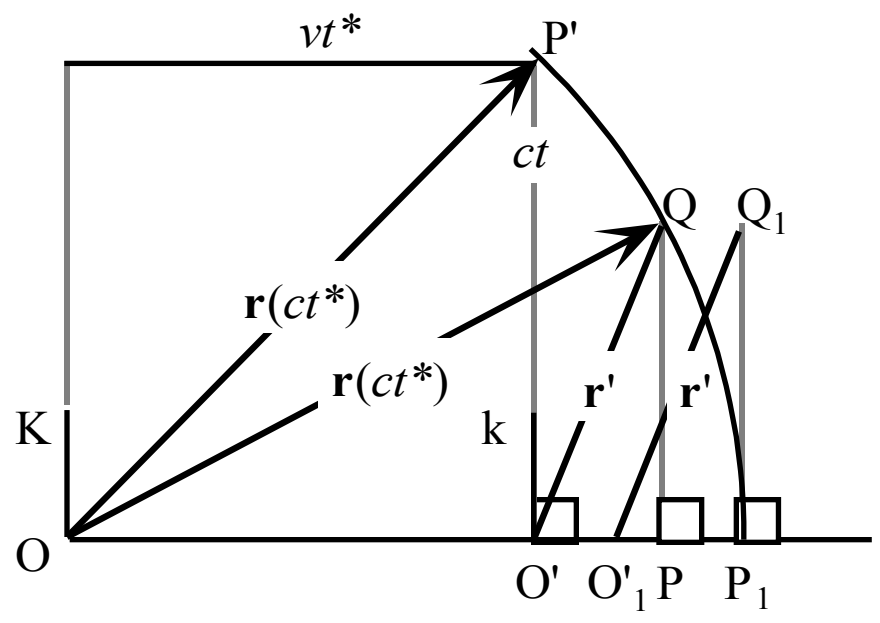

Figure 9.

From a diagram analogous to that in Fig. 9, describing the rectilinear motion of constant velocity $\mathbf{w}$ of a CS k relative to the CSAR K, we obtain analogously the vector LT

$$\mathbf{r}'' = \mathbf{r} - (\mathbf{r}\cdot\hat{\mathbf{w}})\hat{\mathbf{w}} + \gamma[(\mathbf{r}\cdot\hat{\mathbf{w}})\hat{\mathbf{w}} - \mathbf{w}T], \; t'' = \gamma[T - \mathbf{r}\cdot\mathbf{w}/c^2] \qquad (21)$$

where $\hat{\mathbf{w}} = \mathbf{w}/w$, $\gamma = 1/\sqrt{1 - w^2/c^2}$ and $T = \mathbf{r}\cdot\hat{\mathbf{w}}/c$

## Appendix 2: Group Properties

The main mathematical requirement for a set of coordinate transformations to form a group is that they to accomplish the transitivity property. This stipulates that, successively performed, any two of them engender an equivalent one; *i.e.* both collinear and non-collinear LT's form a group. Proving this by the operational method developed in Sect. 3 and Appendix 1 requires tracing of radius vectors by light signals. Note that $O'_{If}$ in Figs. 10 and 11 is the origin of the CSAR $K_A$ associated to $k_A$ as in Sect. 2.1. Tracing $O'_{If}P_{IB}$ and $O'_{If}P_C$ in Figs. 10 and 11, respectively , one finds new transformations related to (20) and (21) and similar to them. The light signals will leave $O'_{If}$ when $O'_{If}$ and the origin of $k_B$ in Fig. 10 (that of $k'_B$ in Fig. 11) coincide. They will reach $P_{IB}$ in Fig. 10 ($P_{If}$, $P_C$ in Fig. 11) simultaneously with the light signal leaving O together with the origins of $k_A$ and $k_B$, when the origin of $k_B$ reaches $O'_{IB}$ in Fig. 10 ($O'_{IB'}$, $O'_{IB}$ in Fig. 11). As concerns the inverse transformation, it is associated with the motion with constant speed $-v$ of the origin of K from O' to O in Fig. 3 relative to the k now at absolute rest. It connects coordinates and times defining a different event. This because the CSAR Ξ associated to the moving K by $\xi = \beta^2 x$ differs from that associated with the moving k by $\xi = \beta^2 x'$ [predicted by (18) in view of (17) and (3)].

## A. For Collinear LT's

Consider the diagram in Fig. 10 for the collinear LT's (20), (21). At $t = 0$ the coinciding origins of $k_A$, $k_B$ and a light signal leave the origin O of the CSAR K. The points $O'_A$, $O'_B$ in Fig. 10 are reached by the origins of $k_A$, $k_B$, respectively, at time $T$, when the light signal reaches P( $X$). In accord with Sect. 3 above, the LT's (20), (21) are written at the times $\beta T$ and $\gamma T$, respectively. The origin of $k_A$ moves from $O'_{IA}$ to $O'_{If}$ in the time $\gamma T - \beta T$. Analogously to the motion of k relative to $K_A$ in Sect. 2.1, we consider the motion of $O'_{IB}$ in relation to $O'_{If}$. From Fig. 10 we have $\mathbf{r}'' = \mathbf{R} - \mathbf{O'_{If}O'_{IB}}$ with

$$O'_{If}O'_{IB} = w\gamma T - v\gamma T,\ (\mathbf{R}\cdot\hat{\mathbf{w}}) = \gamma X - v\gamma T = \gamma(X - vT) = \gamma x'/\beta$$

where $x'$ is just $\xi$ in (14), and

$$\begin{aligned} x'' = (\mathbf{r}''\cdot\hat{\mathbf{w}}) &= \gamma x'/\beta - (w-v)\gamma T = \gamma x'/\beta - \gamma(w-v)\beta t' - \gamma(w-v)\beta v x'/c^2 \\ &= \gamma(x'/\beta)[1-(w-v)v/(c^2-v^2)] - \gamma\beta(w-v)t' \\ &= \gamma\beta(1-wv/c^2)x' - \gamma\beta(w-v)t' \end{aligned}$$

where $t'$ is just $\tau$ in (14).

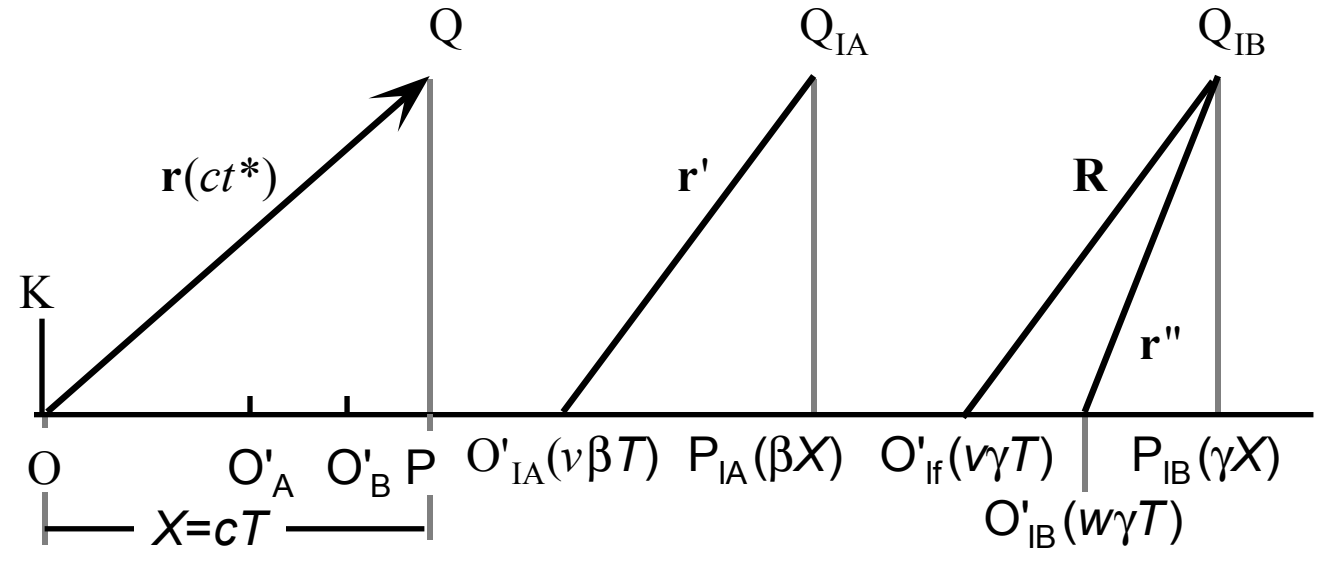

Figure 10.

With $\hat{\mathbf{u}}$ given by (4), $\hat{\mathbf{u}} = \mathbf{u}/u$, $\delta = 1/\sqrt{1-u^2/c^2}$, and $\hat{\mathbf{v}}, \hat{\mathbf{w}}, \hat{\mathbf{u}}$ all parallel, the relationships

$$\gamma\beta(1 - \mathrm{wv}/\mathrm{c}^2) = \delta,\ \gamma\beta(w-v) = \delta u,\ \text{and } \mathbf{r}'\cdot\hat{\mathbf{u}} = x' \tag{22}$$

follow. From the right triangle $O'_{IB}\,Q_{IB}\,P_{IB}$ and the right triangle $O'_{IA}\,Q_{IA}\,P_{IA}$ ($Q_{IA}\,P_{IA}=Q_{IB}\,P_{IB}$), we get the new vector LT

$$\mathbf{r}'' = \mathbf{r}' - (\mathbf{r}'\cdot\hat{\mathbf{u}})\hat{\mathbf{u}} + \delta[(\mathbf{r}'\cdot\hat{\mathbf{u}})\hat{\mathbf{u}} - \mathbf{u}t'],\ \mathrm{t}'' = \delta\,[\mathrm{t}' - \mathbf{r}'\cdot\mathbf{u}/\mathrm{c}^2],$$

where $t' = \mathbf{r}'\cdot\hat{\mathbf{u}}/c$ and $t'' = \mathbf{r}''\cdot\hat{\mathbf{u}}/\mathrm{c}$, which relates position vectors of geometrical points relative to $k_B$ and $k_A$. Thus the transitivity condition is proved for collinear LT's.

## B. For Non-Collinear LT's

Consider the diagram in Fig. 11. At time $t=0$ the CS's $k_A$ and $k_B$, whose origins coincide with that of CSAR K start moving along non-parallel directions with constant velocities $\mathbf{v}$ and $\mathbf{w}$, respectively. Also at time $t=0$, light signals start traveling towards $P_A$ and $P_B$, respectively. To prove that the resulting non-collinear LT's (20), (21) form a group, a light signal and a CS parallel to $k_B$ must move simultaneously at absolute speeds $c$ and $|\mathbf{w}-\mathbf{v}|$ along $O'_AO'_B$ in the time $T$. A new LT, in relationwith (20) and (21) should follow. To this end, we further consider a CS $k'_B$ parallel to $k_B$ which covers in the time $T$ a distance equal to $OO'_A + O'_AO'_B$ along $OP_A$ at a constant velocity $\mathbf{w}^*$. This CS defines a CS $k''_B$, also parallel to $k_B$. The origin of $k''_B$ leaves $O'_A$ at time $t=0$, and, moving with speed $w^*-v$, reaches $O'_B$ at time $T$. So we pass from the relative speed $|\mathbf{w}-\mathbf{v}|$ to the relative speed $w^*-v$ by $|\mathbf{w}-\mathbf{v}|\,T = (w^*-v)T$, and from the motion of $k_B$ relative to $k_A$ to one relative to the CSAR $K_A$, associated to $k_A$ by $(\mathbf{w}^*-\mathbf{v})T = (T - w^*vT/c^2)\,u\,\hat{\mathbf{u}}$ = with

$$u = (w^*-v)/(1-w^*v/c^2) \text{ and } \hat{\mathbf{u}} = (\mathbf{w}-\mathbf{v})/|\mathbf{w}-\mathbf{v}|$$

Using

$$\mathbf{w} - \mathbf{v} = (w^*-v)\hat{\mathbf{u}} \tag{23}$$

we have the operational law for the composition of non-parallel velocities.[3]

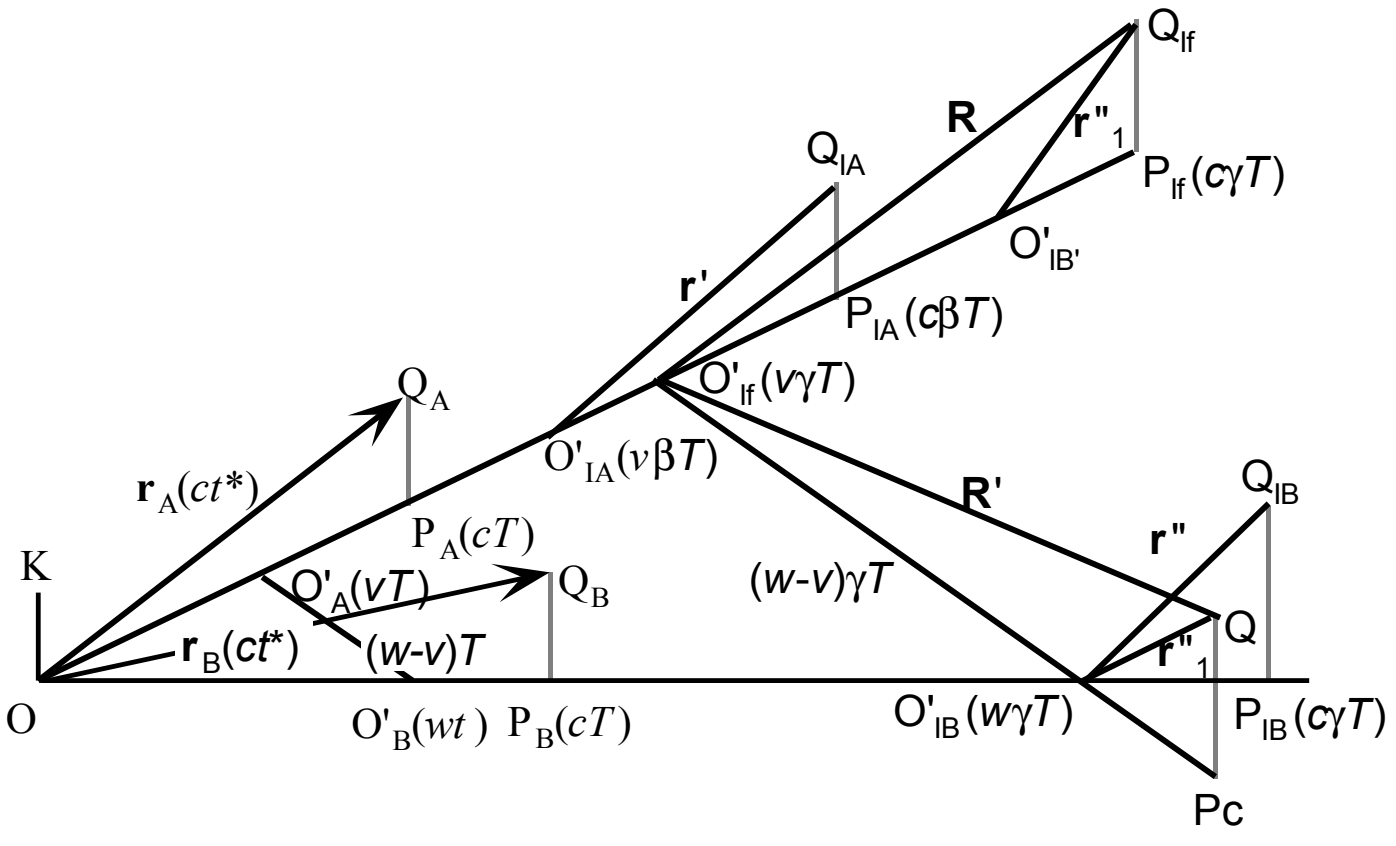

Figure 11.

At the times $\beta T$, $\gamma T$ the light signals that leave O simultaneously with $k_A$, $k_B$ and $k'_B$ reach, respectively, $P_{IA}$ and $P_{If}$, $P_{IB}$. The origins of $k_A$ and $k_B$ arrive, respectively, at $O'_{IA}$ and $O'_{If}$, $O'_{IB}$. In accordance with Sect. 2.1, $O'_{If}$ is the origin of the CSAR $K_A$ at time $\gamma T$. By the above definition of $k'_B$ and $k''_B$, the origin of $k'_B$ finds at time $\gamma T$ at a distance equal to $O'_{If}O'_{IB}$ from $O'_{If}$ along $OP_{If}$, namely at $O'_{IB'}$ in Fig. 11. The light signals leaving $O'_{If}$ simultaneously with the origins of $k'_B$ and $k''_B$ will travel equal distances along the directions of motion of $k'_B$ and $k''_B$, *viz.* $O'_{If}P_{If}=O'_{If}P_C$. Since $O'_{If}P_{If}$ is the projection of $\mathbf{R}$ onto the direction of $\mathbf{v}$, $O'_{If}P_C$ will be the projection of a vector $\mathbf{R'}$ of magnitude $R$ that makes with $\hat{\mathbf{u}}$ an angle equal to that $\mathbf{R}$ makes with $\mathbf{v}$. From $O'_{If}P_{If}=\mathbf{R}\cdot\hat{\mathbf{v}}=\gamma(\mathbf{r}\cdot\hat{\mathbf{v}}-vT)$ and an equation resulting from the first of Eqs. (20), $\mathbf{r'}\cdot\hat{\mathbf{v}}=\beta(\mathbf{r}\cdot\hat{\mathbf{v}}-vT)$, we have $\mathbf{R}\cdot\hat{\mathbf{v}}=(\gamma/\beta)\mathbf{r'}\cdot\hat{\mathbf{v}}$ with

$$(\mathbf{R}\cdot\hat{\mathbf{v}})\hat{\mathbf{u}}=(\gamma/\beta)(\mathbf{r'}\cdot\hat{\mathbf{v}})\hat{\mathbf{u}} \tag{24}$$

By inserting (24), the inverse of the last of Eqs. (20), and Eq. (23) into $(\mathbf{R}\cdot\hat{\mathbf{v}})\hat{\mathbf{u}}-\mathbf{u'}\gamma T$, we obtain:

$$\begin{aligned}
&(\mathbf{R}\cdot\hat{\mathbf{v}})\hat{\mathbf{u}}-(\mathbf{w}-\mathbf{v})\gamma T\\
&=\frac{\gamma}{\beta}(\mathbf{r'}\cdot\hat{\mathbf{v}})\hat{\mathbf{u}}-\hat{\mathbf{u}}(w^*-v)\gamma\beta t'-\frac{1}{c^2}\hat{\mathbf{u}}(w^*-v)\gamma\beta v(\mathbf{r'}\cdot\hat{\mathbf{v}})\\
&=\frac{\gamma}{\beta}(\mathbf{r'}\cdot\hat{\mathbf{v}})\hat{\mathbf{u}}[1-(w^*-v)v/(c^2-v^2)]-\hat{\mathbf{u}}(w^*-v)\gamma\beta t'\\
&=\gamma\beta(1-wv/c^2)(\mathbf{r'}.\hat{\mathbf{v}})\hat{\mathbf{u}}-\hat{\mathbf{u}}(w^*-v)\gamma\beta t'
\end{aligned}$$

In view of Eqs. (24), also valid for $w^*$, we have:

$$\mathbf{O'_{IB}P_C}=(\mathbf{R}\cdot\hat{\mathbf{v}})\hat{\mathbf{u}}-(\mathbf{w}-\mathbf{v})\gamma T=\delta[(\mathbf{r'}\cdot\hat{\mathbf{v}})-ut']\hat{\mathbf{u}}=\delta[(\mathbf{r'}\cdot\hat{\mathbf{u}})\hat{\mathbf{u}}-\mathbf{u}t']$$

Because $Q_{If}P_{If}=Q_{IB}P_{IB}=QP_C$ by virtue of $Q_AP_A=Q_BP_B$, and $|\mathbf{r}_1''|=|O'_{IB'}Q_{If}|=|O'_{IB}Q|$ with $\mathbf{O'_{IB}Q}=\mathbf{QP_C}+\mathbf{O'_{IB}P_C}$, we have $\mathbf{QP_C}=\mathbf{r'}-(\mathbf{r'}\cdot\hat{\mathbf{u}})\hat{\mathbf{u}}$ and

[3] This law has no physical grounding in common with the standard relativistic formula for the composition of non-parallel speeds [19] -which predicted the famous, but contested [20] Thomas precession. For the sake of mathematical generality, Thomas missed the physical meaning of the LT by the translation he associated to the vector LT [9]. It was under such condition that the usual matrix multiplication he used to made no physical sense.

$$\mathbf{r''}_1 = \mathbf{r'} - (\mathbf{r'}\cdot\hat{\mathbf{u}})\hat{\mathbf{u}} + \delta\,[(\mathbf{r'}\cdot\hat{\mathbf{u}})\hat{\mathbf{u}} - \mathbf{u}t'\,],\;\; t'' = \delta[\,t' - \mathbf{r'}\cdot\mathbf{u}\,/\,c^2\,] \qquad (25)$$

where $t'' = \mathbf{r''}_1 \cdot \hat{\mathbf{u}} / c = \mathbf{r''}_1 \cdot \hat{\mathbf{w}} / c$. The resulting vector LT (25) proves that the non-collinear LT's satisfy the transitivity property. Hence they form a group without requiring rotations of 'stationary' (inertial) CS's in this aim. This result validates LT itself.

## Appendix 3: Outline of the Crisis of Modern Physics

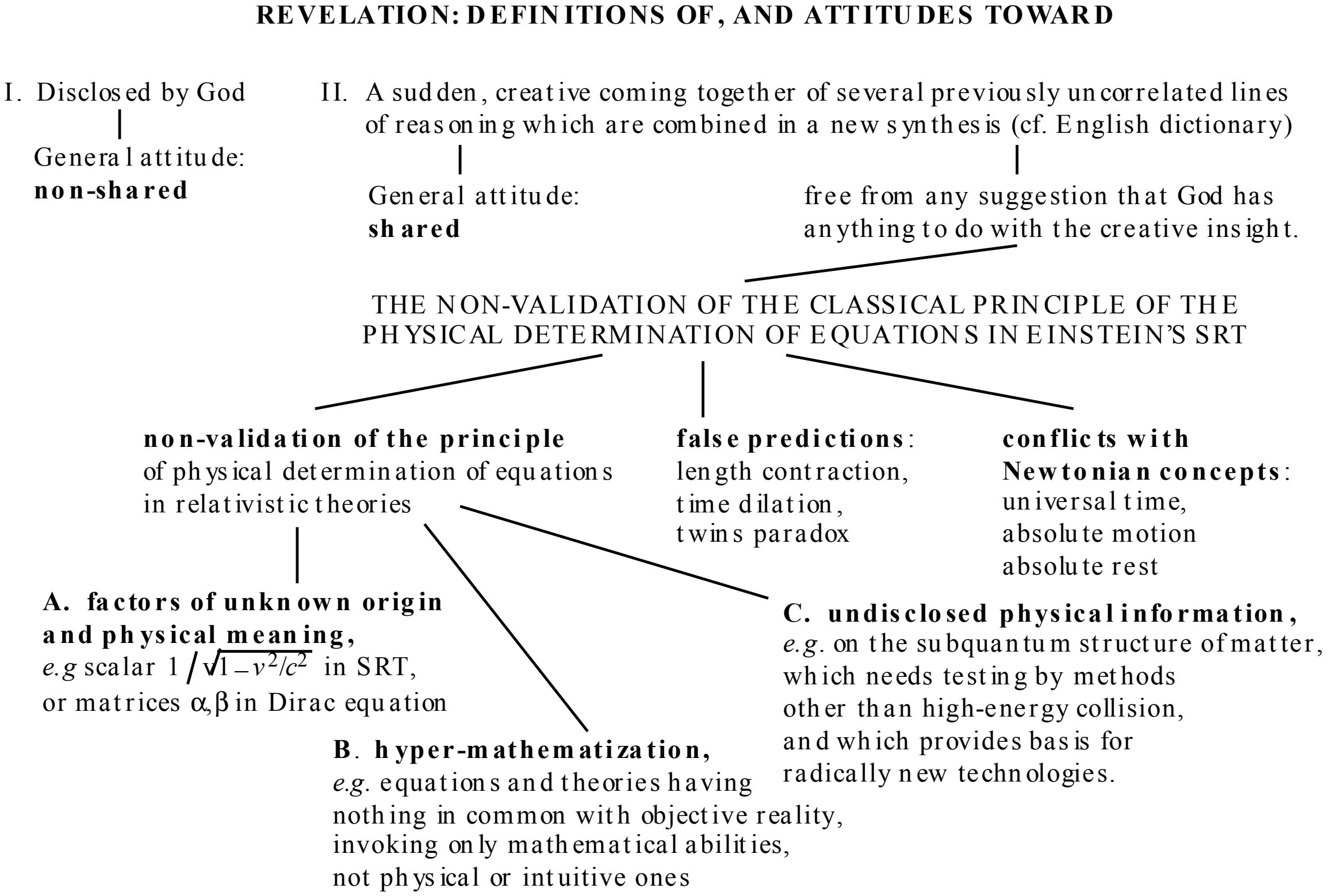

**A+B+C LED TO THE CRISIS IN MODERN PHYSICS**

**Defining features:** much high mathematics, little physical information

**Development:** first perceived: around the 1940's; really started: 1905

**Root cause:** physicists attitude concerning the role of revelation in 'the act of science'; *i.e.*, the birth of any new idea, or set of coupled ideas, contributing to the advancement of science.

## Appendix 3: Revelation's Role in the Act of Science

A successful trend of both science and secularization accredited the idea that science and divine work would be antinomies. Physicists supported this idea by *a fortiori* interpreting failures in the theoretical work as natural steps toward the truth, so disregarding -against the evidences- the century-old crisis of physics. No role is granted to revelation (as disclosure by God) in the act of science. A definition of revelation free from any suggestion that God has anything to do with the creative insight was put forward as "a sudden, creative coming together of several previously unconnected lines of reasoning which are combined in a new synthesis" (English dictionary). When faced up to the "incomprehensible" successful work of some among them, "who did not seem to be reasoning at all but who jumped over all intermediate steps to a new insight about nature" [21], physicists confined to name them "magicians", and 'felt' "compelled to redo the work of the magicians so that they seem like sages" [21] ("sages" are those physicists who "reason in an orderly way about physical problems on the basis of fundamental ideas of the way that nature ought to be" [22]). They claimed that "otherwise no reader would understand the physics" [21]. Then they established a 'prophylactic' editorial quarantine against new "magicians".

This is the mainstream in modern physics. In despite its strategy, the crisis (see Appendix 3) is evolving. It means that something is wrong with this strategy. Whether discarding any role to revelation in the act of science seemed to be a natural attitude when physics emancipated as science by measurements and elementary mathematics, it became questionable when syntheses of experimental data, novel ideas and advanced mathematics faced physics. To resolve the dilemma, a question is essential to be answered: Whether revelation (as disclosure by God) plaid indeed a role in the act of science, could its mark be identified in the valuable works of the physicists denying its role, or just believing (like Einstein) that a revealed knowledge cannot be rationalized? To this end, consider our results in Sect. 6, concerning the derivation of the LT in [2]. We conclude that Einstein was playing the role of a "magician" -the most important- because:

First, he "jumped over all intermediate steps" -consisting in the physical motivation of the manipulations of equations that led to the LT. Our derivation of the CT's, and of the LT as a CT enabled us to disclose (Sect. 6) the objective reality warranting his manipulations of equations (see the three mathematical decisions pointed out in Sect. 5). This objective reality consisted in tracing the radius vectors with light signals. Hence, in despite their strong appearance of mathematical tricks, the manipulations were not tricks at all. The derivation of the LT in [2] was correct.

Second, he "did not seem to be reasoning at all". He discarded the concepts of absolute rest and absolute motion but described in detail a thought experiment which seems to be the only one enabling the 'blind' inertial observers to determine absolute speeds in their reference frames (see the diagram of this experiment in Sect. 5). He proposed the experiment for deducing the LT in the idea that identical inertial clocks would run at rates depending on their speed. But, because he did not realize the role plaid by the light signals in this experiment, needed to manipulate some equations (pointed out in Sect. 5) to this end. Unfortunately, he did not investigate further the diagram describing the experiment to see that this diagram actually validates 'abstract' CSAR's in his theory (as we proved in Sect. 6).

There becomes evident that Einstein was not aware that i) by light signals has specified the time-changing magnitude and direction of the radius vectors of geometrical points moving with respect to inertial observers (which should lead him to the LT as a CT) but he used light signals, ii) the graphical addition of travel times as scalar quantities (which we developed in Sect. 3) needed be developed in his theory but he worked only with light signals tracing abscissas of geometrical points and dropped the square of $\beta$ in his equations linear in $\beta^2$, according to the graphical addition of travel times as scalar quantities, iii) the equation $x = wt$ assured the independence of the linear equations in $\beta$ (making them a coordinate transformation) but he took into account this equation in order to obtain the "addition theorem for speeds" [2] (Sect. I.5) and iv) the CSAR plays an essential role in his theory but he consecrated a version of the light-speed principle in [2] (Sect. I.2) that saved his theory from the inconsistencies raised by the arbitrary removal of the CSAR.

It is as if Einstein reconstituted by flashes in [2] a paper on the derivation of the LT as a CT that pre-existed in his subconscious. The correctness of all the manipulations of equations (the clue of [2]) supports the revealed nature of the original paper. The lack of their physical motivation shows that Einstein turned into rational knowledge only pieces of revealed knowledge. That is why he never became aware of the correctness of the derivation of the LT in [2], and, fatally, developed special relativity theory without the derivation of the LT in [2].

Einstein's correct derivation of the LT in [2], and his later disregarding of it are the most striking proof that revelation plays an essential role in the act of science. Once we identified the mark of revelation in [2], it is (more or less) identifiable in the valuable work of any physicist. Unfortunately, when it happened, the identification of the mark of revelation was not followed by a rationale of the work. The "jumps over the intermediate steps" of the authors were not filled with the missed information. The work identified as revealed (like [2]) became thereafter *unalterable*, of *eternal value*, completely foreign to the advancement of physics. The identification of the mark of revelation by authors themselves depends on their attitude toward revelation. The discarding of the revelation role in the act of science allows physicists to take rational decisions which strongly disturb their revealed knowledge. So are raised the "jumps over intermediate steps" -particularly of explanatory nature- in their work, the loosing and distortedly perceiving of essential physical information. The crisis of modern physics is the result of disregarding all these evidences. It is the unseen, dark face of the secularization. So fundamental for the eradication of this crisis is the physicists' acceptation that revelation plays certainly the key role in the act of science.

Far -by his development of special relativity theory without the derivation of the LT in [2], and the foundation of modern physics on SRT- Einstein was the main contributor to the crisis of modern physics. Other key contributors were the great physicists P.A.M. Dirac and B.L. van der Waerden (who disregarded revelation). Both they missed the subquantum information embedded in Dirac's equation. Actually, like Einstein, they failed in rendering conscious the whole information revealed them through their subconscious (*humans touch divine through their subconscious*). Their work stands for proof that they couldnot provide a complete rationale for the revealed knowledge. They, like all the "magician-physicists", behaved as if have had accessed intermittently a superhuman database.

As to the impact of the missed revealed knowledge on the human progress, let us examine the consequences of the derivation of the LT in [2], if Einstein gave a complete rationale for it. Most important, he obtained (without exception) the terms of the LT as Cartesian coordinates and Newtonian times. There became evident the lack of any conflict between Einstein's and Newton's theories. The principle of the physical determination of equations worked successfully in both theories. There has been no mental alienation by the famous time dilation and twin paradox. The validation of the principle of the physical determination of equations in modern theories concerning the quantum and subquantum structure of matter through the relativistic energy-momentum relationship should follow. Dirac and der Waerden should obtain genuine subquantum information. The application of this information to radically new technologies (that should happen as early as by the 1940's) gives the real dimension of the impact which the missed and distortedly perceived revealed knowledge had (still has) upon the advancement of physics, finally upon the progress of the mankind.

However, decoding the revealed knowledge is not so easy. Einstein's failure in providing a rationale for the derivation of the LT in [2] points to the existence of some hardly to identify but easily 'deletable' passwords for accessing the understanding of a revealed knowledge. The concepts of absolute rest and absolute speed prove to have been such 'passwords'. These 'paswords' were 'deleted' neither by Einstein's followers nor by Einstein after ending SRT but by Einstein in the preamble of his original paper on relativity [2], when stated that "no properties of phenomena attach to the idea of absolute rest". So that an undisturbed conversion of a revealed knowledge into a rational one is assured by a careful search for hidden passwords and a careful choise of decisions. Discarding or disregarding the role plaid by revelation in the act of science, so these requirements, substantially affects physicists' performance. Breaking (like individuals) the atheistic mentality (beneficiary of a formidable logistics), as well as the mentality that revealed knowledge cannot be turned into rational knowledge is needed to this end.

The rationale which we give for the first time to a revealed knowledge suggests that people can access some revealed knowledge benefic to the material progress of the mankind.

## References[4]

[1] A. Einstein, Out of my Later Years, pp. 53, 41, 40 (Wings Books, New York, Avenel, New Jersey, 1993).

[2] A. Einstein, "Zur Elektrodynamik bewegter korper", Annalen der Physik 17, 891 (1905).

[3] A. Einstein, "Uber das Relativitatsprinzip und die aus demselben gezogene Folgerungen", Jahrbuch der Radioaktivitat und Electronik 4, 411 (1907).

[4] J.A. Wheeler, Geometrodynamics (NY, Acad. Press, 1962).

[5] E. Taylor and J.A. Wheeler, Spacetime Physics (W.H. Freeman and Co, San Francisco, London, 1966).

[6] C.W. Misner, K.S. Thorne and J.A. Wheeler, Gravitation (W.H. Freeman and Co, San Francisco, London, 1973).

[7] S. Weinberg, "A Unified Physics by 2050?", Sci. Am. 281, 68 (1999).

[8] A.C.V. Ceapa, Physical Grounds of Einstein's Theory of Relativity; Roots of the Falsification of 20th Century Physics, 3rd ed., part III (Bucharest, l998, LC call No QC173.55.C43 1998 ).

[9] L.H. Thomas, "The Kinematics of an Electron with an Axis", Phil. Mag. (7), 3, 1 (1927).

[10] A.C.V. Ceapa, "Full Physical Derivation and Meaning of Lorentz transformation", http://arxiv.org/abs/physics/9911067.

[11] A. Einstein, Out of my Later Years, p. 39 (Wings Books, New York, Avenel, New Jersey, 1993).

[12] J.C. Hafele and R.E. Keating, "About the World Atomic Clocks: Predicted Relativistic Time Gains", Science 177, 166 (1972).

[13] A.G. Kelly, "Reliability of Relativistic Effect Tests on Airborne Clocks", Monograph 3 (Institute Engineering Ireland, 1996 ).

[14] D. Saa, "Frequent Errors in Special Relativity", http://arxiv.org/abs/physics/0506207.

[15] S.J. Prokhovnik, The Logic of Special Relativity, p. 89 (Cambridge University Press, l967).

[16] A.C.V.Ceapa,"Coordinate Transformations Between Coordinate Systems in Relative Motion", Phys. Essays 4, 60 (1991).

[17] A.C.V. Ceapa, "Relativistic Theories like Operational Theories", in Abstracts of Contributed Papers. 12th Internatl. Conf. On General Relativity and Gravitation, p. 158 (Boulder, 1986).

[18] L.D. Landau and E.M. Lifshitz, The Classical Theory of Fields, Ch. 101 (Pergamon, N.Y., 1980).

[19] A.A. Ungar, "The Relativistic Velocity Composition Paradox and the Thomas Rotation", Found. Phys. 19, 1385 (1989).

[20] T.E. Phipps, Jr., "Kinematics of a Rigid Rotor", Lett. Nuovo Cim. 9, 467 (1974).

[21] S. Weinberg, Dreams of a Final Theory (Vintage Books, Random House, Inc., New York, 1994) p. 68.

[22] Ibid., p. 67.

[4] A. Ceapa and A.C.V. Ceapa stand in these references for one and the same author.